\documentclass[aca4]{JAC2000}

\usepackage{graphicx}

\begin{document}
\title{Double Kicker system in ATF}

\author{T.Imai, K.Nakai, Science University of Tokyo, Chiba, Japan\\
H.Hayano, J.Urakawa, N.Terunuma, KEK, Ibaraki, Japan}

\maketitle

\begin{abstract} 
A double kicker system which extracts the ultra-low emittance multi-bunch
beam stably from ATF damping ring was developed.
The performance of the system was studied comparing an orbit jitter with
 single kicker extraction in single bunch mode.
The position jitter reduction was estimated from the analysis of the extraction
 orbits. 
The reduction was confirmed for the
 double kicker system within a resolution of BPMs. 
More precise tuning
 of the system with a wire scanner has been tried by changing a $\beta$ function at the
 second kicker to get more reduction of kick angle jitter.
The results of these studies are described in detail. 
\end{abstract}

\section{introduction}

   KEK/ATF is an accelerator test facility for an injector part of a future
   linear collider. It consists of an S-band injector
   linac, a beam-transport line, damping ring and extraction line
   \cite{atf_design}. 
   The main purpose of ATF is to generate and measure ultra-low
   emittance multi-bunch beam (2.8nsec spacing, 20bunch) and develop 
   technology that can stably
   supply the beam to the main linac.
 
\begin{figure}[htbp]
\centering
\includegraphics*[width=75mm]{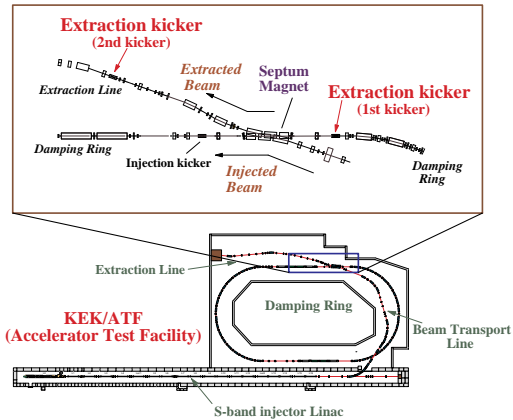}
\caption{Layout of injection/extraction region of ATF}
\end{figure}

   The stable beam extraction from the damping ring is essential for linear
   collider to achieve high luminosity, because the position
   jitter would be magnified by transverse wakefields in the
   linac and reduce the luminosity. Therefore, the jitter tolerance of extraction kicker magnet is
   very tight and estimated to be $5 \times 10^{-4} $ assuming
   $\beta_{x}$ = 10m\cite{TOR}. 
   It will be applied not only to a 
   uniformity of pulse magnetic field for the tolerance of multi-bunch
   but to a pulse-to-pulse stability. In ATF, double kicker system was developed for the stable beam
   extraction \cite{DK2}. 
   The system uses two identical kicker magnets for beam extraction.
   The first kicker is placed in the damping ring and the second one
   for jitter compensation in the extraction line.


The ATF extraction kicker consists of 25 electrode pairs with ferrite
loaded in a vacuum chamber. A ceramic tube with TiN coated inside is
used for beam channel in the kicker in order to reduce an beam
impedance. The specification of the kicker is summarized in Table~\ref{kicker_spec}. 

\begin{table}[htb]
\begin{center}
\caption{ATF Extraction Kicker magnet}
\begin{tabular}{lll}                 \hline\hline
  Kick angle            & 5          & mrad        \\
  Impedance             & 50         & $\Omega$    \\ 
  Magnet length         & 0.50       & m           \\
  Magnetic field        & 513        & Gauss       \\
  Rise and Fall time    & 60         & nsec        \\
  Flat top              & 60         & nsec        \\
  Maximum voltage       & 40         & kV          \\
  Maximum current       & 800        & A           \\\hline\hline
\end{tabular}
\label{kicker_spec}
\end{center}
\end{table}

\section{double kicker system}

The double kicker system consists of one pulse power supply and two
kicker magnets separated by phase advance $\pi$. It is, in principle, 
able to compensate kick angle variation of the extraction kicker in
damping ring.
When each kicker has a kick angle variation $\Delta \theta_{1}$ and $\Delta
\theta_{2}$, $(x,x^{\prime})$ at the second kicker can be written as
\begin{equation}
{x \choose x^{\prime}} =  M_ {1 \rightarrow 2}{0 \choose \Delta \theta_{1}} + {0 \choose \Delta \theta_{2}} \label{eq1}
\end{equation}
Here, $M_ {1 \rightarrow 2}$ is a transfer matrix from the first kicker
to the second one. Since
a phase difference of the two kickers is $\pi$,
\begin{equation}
   {x \choose x^{\prime}} = 
    \pmatrix{
      -\sqrt{\frac{\beta_{2}}{\beta_{1}}}                     &  0                                  \cr
      -\frac{\alpha_{2}-\alpha_{1}}{\sqrt{\beta_{2}\beta_{1}}} & -\sqrt{\frac{\beta_{1}}{\beta_{2}}} \cr
             }{0 \choose \Delta \theta_{1}}+{0 \choose \Delta \theta_{2}} \label{eq2}
\end{equation}
then,
\begin{equation} 
  x = 0 ,     x^{\prime} =   -\sqrt{\frac{\beta_{1}}{\beta_{2}}} \Delta \theta_{1} + \Delta \theta_{2}
\end{equation}
are obtained.
   When the two kickers are identical, that is $\Delta \theta_{1} = \Delta
   \theta_{2} $, the variation could be canceled with the same $\beta$
   function. 
If the two kickers are not
   identical, compensation also can be done by
   adjusting $\beta$ function. 
In case of multi-bunch extraction, only the similarity of the flat-top
waveform between two kickers is required for the jitter reduction in each
bunch. A tight flatness requirement for every bunch in a multi-bunch is
not necessary.
\begin{figure}[htbp]
\centering
\includegraphics*[width=75mm]{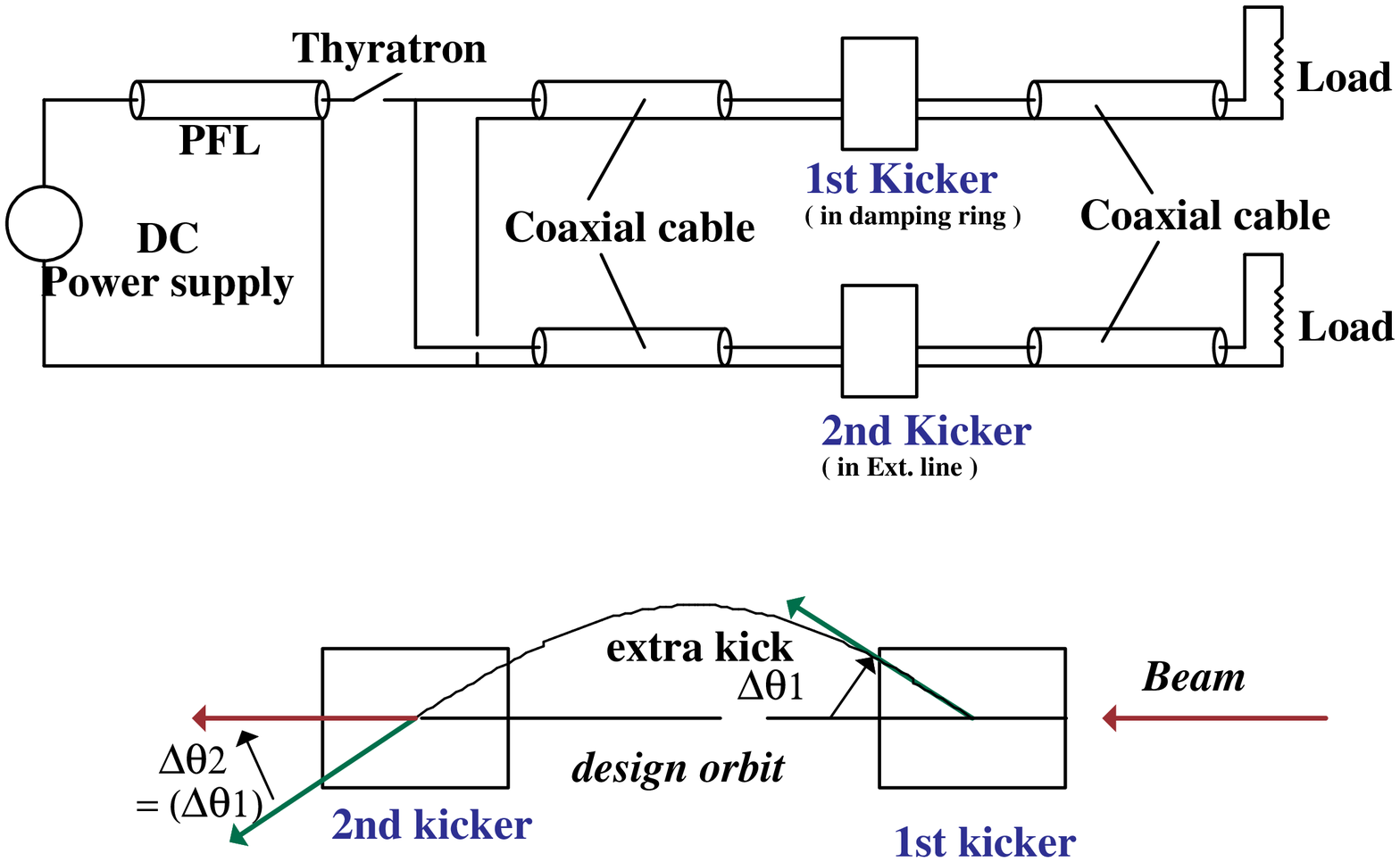}
\caption{Double kicker system}
\end{figure}

\section{orbit jitter reduction}

We measured horizontal beam orbit jitter of the extraction line in single
bunch operation and compared the performance of the double kicker system 
with the case of extraction without the second kicker which we defined as a single kicker mode.

The operation condition at the measurement was single bunch mode, beam
energy 1.3GeV, and repetition rate is 0.78Hz. 
The beam is extracted from damping ring to the extraction line by one kicker
magnet and three DC septum magnets. 
Kick angle of the kicker is designed to 5mrad. 
In single kicker mode, a dipole magnet was installed instead the second kicker.

The beam orbit shot by shot was measured by strip-line BPMs in the extraction line
\cite{BPM}. 
Total 14 BPMs are used for the measurement and analysis.

\subsection{Analysis method}

The horizontal beam position jitter in the extraction line came from a kick angle jitter of
the extraction kickers and a momentum fluctuation of the beam. The horizontal
displacement at a point after the second kicker can be written as

\begin{eqnarray}
\Delta x_{i}&=& \Delta x_{kicker} + \Delta x_{momentum} \nonumber \\
            &=& R_{12}(1,i) \Delta \theta_{1} +  R_{12}(2,i) \Delta \theta_{2} + \eta_{i} 
\frac{\Delta p}{p}
\end{eqnarray}
\label{eq4}

where $R_{12}(1,i)$ is a transfer matrix component from the first kicker
to the point i , $R_{12}(2,i)$ is from the second one and $\eta_{i}$ is
a dispersion.
In this analysis,
$\Delta \theta_{1}$,$\Delta \theta_{2}$ and  $\Delta p /p$ were obtained
by fitting the measured displacement from the average position at BPMs with eq(4).
$\Delta x_{kicker}$ was calculated by subtraction
$\eta_{i} \frac{\Delta p}{p}\label{eq4}$ from the displacement at each BPM.
In single kicker mode, the same analysis also has been done, 
but $\Delta \theta_{2}$ was set to zero 
because the dipole magnet installed as the second kicker is DC magnet.


\subsection{Result of jitter reduction}

   Table~\ref{mode_res} shows the comparison of the position jitter caused by kick angle
   variation in both modes at the BPM where has the maximum $R_{12}$ from
   the kicker.

$\sigma_{kicker}$ was calculated by using model value of
$R_{12}$ and fitting value of kick angle variation of each shot. 
The effect of jitter reduction was compared in 
the two kick angle region. 
As a result,
the double kicker system reduced the jitter down to the resolution of BPM which
is about 20$\mu$m in case of small kick variation, 
however the reduction rate was not sufficient in case of large kick variation. 
In both cases, the position jitter reduction was observed for the double
kicker configuration, however, a precise tuning of the system and a high
resolution position monitor are still necessary for a further reduction.

\begin{table}[htb]
\begin{center}
\caption{Comparison of position jitter measurement}
 \begin{tabular}{c|cc|cc}                 \hline
mode&$\#$ of&$\sigma_{kicker}$&$\#$ of&$\sigma_{kicker}$\\ 
& meas.&[$\mu$m] & meas.&[$\mu$m] \\\hline
double&115&37&181&24\\  
single&60&78&248&34 \\ \hline
   &\multicolumn{2}{c|}{$\Delta \theta_{1}\ge$0.007mrad} &\multicolumn{2}{c}{$\Delta \theta_{1}\le$0.007mrad}\\ 
 \end{tabular}
\label{mode_res}
\end{center}
\end{table}

\section{optics tuning}

In order to get more jitter reduction, $\beta$ function at the second
kicker was surveyed using high resolution position monitor. One of the
wire scanners was used as a position jitter detector \cite{ws}.

\subsection{Jitter measurement with wire scanner}

The wire scanner which have 10 $\mu$m diameter
tungsten wire was used for the jitter measurement. 
Scattered gamma rays from the wire are detected by air Cherenkov detector with photo
multiplier. 
Before orbit measurement we measured horizontal beam profile and set the
wire at the position which is middle of the slope of the profile. 
The distribution of detected gamma rays is converted to the distribution
of the position with beam profile. 
Horizontal beam size was around 100 $\mu$m in these measurements.

\begin{figure}[hbtp]
\centering
\includegraphics*[width=65mm]{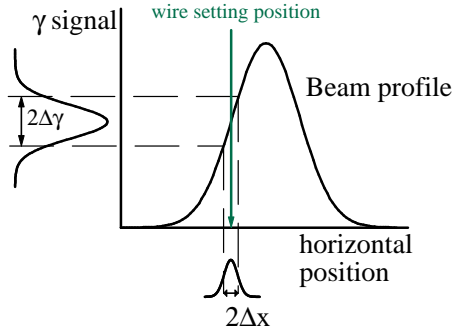}
\caption{Beam jitter measurement using wire scanner}
\end{figure}

\subsection{Optics tuning method}

Changing $\beta$ function at the second kicker, 
a minimum position jitter was surveyed.
However, the condition of jitter compensation 
was different in each optics setting
because dispersion and orbit was corrected in each setting independently. 
Then the position jitter caused by kick angle variation was compared 
by normalizing each condition with  
$\beta$ function 2.5m at the wire and phase advance 1.5 $\pi$ from
the second kicker.  
The result is summarized in Table~\ref{scan_res} . 

\begin{table}[htb]
\begin{center}
\caption{Result of optics tuning}
\begin{tabular}{lllll}                 
\hline
  $\beta$ function & estimated & $\beta$
  function & phase   \\
  2nd Kicker&$\sigma_{kicker}$&wire&advance \\\hline\hline
  5.0 [m]& 11.4 [$\mu$m]& 2.47 [m]& 1.507 [$\pi$] \\
  6.0 & 11.0 & 2.49 & 1.501 \\
  8.0 & 10.9 & 2.65 & 1.492 \\
  10.0 &11.4 & 2.59 & 1.482 \\
  12.0 &12.9 & 1.81 & 1.501 \\\hline
\end{tabular}
\label{scan_res}
\end{center}
\end{table}

With these values 
a $\beta$ function of maximum jitter reduction and 
kick angle jitter are estimated.
Assuming model optics and kick angle ratio $k$,
kick angle variation will be $\Delta \theta_{1}$  = $ k
\Delta \theta_{2}$, horizontal displacement is written by
$\Delta x_{kicker} = R_{12}(1,wire) \Delta \theta_{1}+ R_{12}(2,wire) \Delta \theta_{2}
= \sqrt{\beta_{wire}} \Delta \theta_{1}(\sqrt{\beta_{1}} - k \sqrt{\beta_{2}})$.

By fitting 
\begin{eqnarray}
 \sigma_{estimated} = \sqrt{\sigma_{kicker}^2 +\sigma_{resolution}^2 } \nonumber
\end{eqnarray}
at each optics setting,
\begin{eqnarray}
k &=& 0.833 \nonumber \\
\Delta\theta_{1} &=& 6.7     \mu rad   \nonumber \\
\sigma_{resolution}&=& 10.7   \mu m  \nonumber
\end{eqnarray}
were obtained.

Measured $\beta$ function at the first kicker
was 4.95m, 
so  $\beta$ function at the second kicker for maximum jitter 
reduction was estimated to be 7.13m. 
The estimated resolution 10.7$\mu$m is included
monitor resolution and incoming position and angle jitter.
The resolution of this position monitor using wire is estimated about 1$\mu$m,
then it seemed 
that the beam orbit of damping ring had fluctuation in these measurements. 
The kick angle ratio 0.833 is not explained 
by the difference of the cable length between the two kickers (9.4m).
It seemed that the difference of ceramic coating between the two kickers
caused the much difference of the field strength.

\begin{figure}[hbtp]
\centering
\includegraphics*[width=60mm]{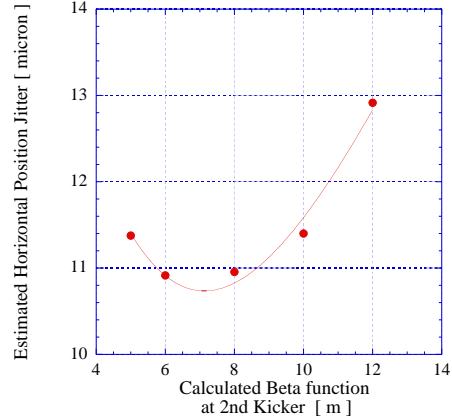}
\caption{Position jitter dependence on $\beta$ function at the second kicker}
\end{figure}

\section{conclusion}

The performance of the double kicker system was studied
with measurement of horizontal orbit jitter in single bunch mode.
The system has the jitter less than the resolution of BPM in small kick variation.
The survey of the maximum reduction optics changing $\beta$
function at the second kicker was performed with wire scanner. 
The $\beta$ function which gave a minimum position jitter was found. 
Estimated $\beta$ function for it was 7.13m 
and this corresponded to the kick angle ratio of the two kickers 0.833.

\section{ACKNOWLEDGMENTS}

The authors would like to acknowledge Professors H.Sugawara, M.Kihara,
S.Iwata and K.Takata for their support and encouragement. We also express our
thanks to all the members of the ATF group for their support to the beam
experiments.


\begin{thebibliography}{9}
\bibitem{atf_design}
Edited by F.Hinode et al.,''ATF Design and Study Report'',KEK Internal 95-4,1995 
\bibitem{TOR}
T.O.Raubenheimer et al.,''Damping Ring Designs for a TeV Linear
	Collider'',SLAC-PUB-4808,1988
\bibitem{DK2}
 H.Nakayama, KEK Proceedings 92-6,1992,p326-p334
\bibitem{BPM}
H.Hayano et al., ``Submicron Beam Position Monitors for Japan
Linear Collider'', LINAC'92,Ottawa,1992
\bibitem{ws}
H.Hayano,''Wire Scanners for Small Emittance Beam Measurement in ATF'',
this conference
\end{thebibliography}
\end{document}